\documentclass[aps,prb,twocolumn,groupedaddress,floatfix]{revtex4-1}
\usepackage{lipsum}
\usepackage{graphicx}
\usepackage{dcolumn}
\usepackage{bm}
\usepackage{amsfonts}
\usepackage{amssymb}
\usepackage{amsmath}
\usepackage{graphicx}% Include figure files
\usepackage{bm}% bold math
\usepackage{color}% bold math

\begin{document}

\title{Collective modes in multi-Weyl semimetals}
%\title{Effect of chirality on collective modes in multi-Weyl semimetals}
%\title{Long wavelength plasmons in three dimensional multi-Weyl semimetals}

\author{Seongjin Ahn$^{1}$}
\author{E. H. Hwang$^{2}$}
\email{euyheon@skku.edu}
\author{Hongki Min$^{1}$}
\email{hmin@snu.ac.kr}
\affiliation{$^1$ Department of Physics and Astronomy and Center for Theoretical Physics, Seoul National University, Seoul 08826, Korea}
\affiliation{$^2$ SKKU Advanced Institute of Nanotechnology and Department of Physics, Sungkyunkwan University, Suwon, 440-746, Korea}

\date{\today}
\begin{abstract}
We investigate collective modes in three dimensional (3D) gapless 
multi-Weyl semimetals with anisotropic energy band dispersions (i.e., $E\sim \sqrt{ k_{\parallel}^{2J} + k_z^2}$ with a positive integer $J$). For comparison, we also consider the gapless semimetals with the isotropic band dispersions (i.e. $E\sim k^N$). We calculate analytically long-wavelength plasma frequencies incorporating interband transitions and chiral properties of carriers. For both the isotropic and anisotropic cases, we find that interband transitions and chirality lead to the depolarization shift of plasma frequencies. For the isotropic parabolic band dispersion the long-wavelength plasmons do not decay via Landau damping, while for the higher-order band dispersions the long-wavelength plasmons experience damping below a critical density. For systems with the anisotropic dispersion the density dependence of the long-wavelength plasma frequency along the direction of non-linear dispersion behaves like that of the isotropic linear band model, while along the direction of linear dispersion it behaves like that of the isotropic non-linear model. Plasmons along both directions remain undamped over a broad range of densities due to the chirality induced depolarization shift. Our results provide a comprehensive picture of how band dispersion and chirality affect plasmon behaviors in 3D gapless chiral systems with the arbitrary band dispersion.
\end{abstract}
\maketitle

\section{Introduction}

In recent years, there have been much interest in Dirac materials such as graphene\cite{CastroNeto2009,DasSarma2011}, topological insulators\cite{Hasan2010,Qi2011}, high-temperature $d$-wave superconductors\cite{Lee2006}, which are described by the Dirac-like equation at low energies. 
Dirac materials show novel physical features because of the chiral band dispersion with spin-momentum locking and the existence of the Weyl points at which conduction and valence bands touch, distinguishing themselves from other conventional materials with parabolic energy dispersion at low energies. 
Recently three dimensional (3D) Weyl semimetals, which are 3D analogs of graphene, have attracted considerable attention because they are expected to give a broad spectrum of unusual physical properties as two dimensional graphene. 
In recent experiments of angle-resolved photoemission spectroscopy and scanning tunneling microscopy, several materials such
as Cd$_3$As$_2$\cite{Neupane2014,Jeon2014,Liu2014,Borisenko2014,He2014,Li2015,Moll2015}, Na$_3$Bi\cite{Liu2014a}, NbAs\cite{Xu2015a},  TaP\cite{Xu2015, Xu2016}, ZrTe$_5$\cite{Chen2015, Li2016}  and TaAs\cite{Lv2015,Lv2015a,Lv2015b,Huang2015a,Xu2015b,Yang2015,Inoue2016} have been identified as Weyl semimetals.
In addition, several efforts have been made on the realization of Weyl semimetals in artificial systems such as photonic crystals\cite{Chen2015a,Zhang2015,Dubcek2015,Lu2015,Lu2016}.

Recently, a new type of Weyl semimetals referred to as multi-Weyl semimetals has been proposed\cite{Xu2011,Fang2012,Huang2016}. These materials are characterized by double (triple) Weyl-nodes in which the band dispersion is 
linear along one direction and quadratic (cubic) along the remaining two directions.
%quadratic (cubic) in the in-plane direction and linear in the out-of-plane direction. 
Because of the change in topological nature and the enhancement of the density of states, the 
anisotropic non-linear energy dispersion and a modified spin-momentum locking structure have important consequences in topological, transport and interaction-induced properties\cite{Banerjee2012,Lai2015,Jian2015,Guan2015,Huang2015, Pyatkovskiy2016,Chen2016,DasSarma2015,DasSarma2015a}. 
Despite their broad interest, however, there have been lack of studies on the collective modes of charge oscillations in chiral gapless systems including multi-Weyl systems. 

The goal of this paper is to investigate electronic collective modes of chiral gapless electron-hole systems in 3D 
and to find wave-vector-dependent plasmon dispersions, considering interband transitions along with chirality of the systems.
Given the great current interest in gapless linearly dispersing Dirac systems and the emergence of multi-Weyl semimetals, it is necessary to obtain results for linear and higher-order dispersions in 
3D in order to develop intuition about the dispersion and chirality dependence of electronic properties.
We consider both gapless semimetals with an isotropic band dispersion and multi-Weyl semimetals with an anisotropic band dispersion. We calculate the plasmon modes and the energy loss functions (spectral strength) in gapless semimetals, and present analytical and numerical results for long-wavelength plasma frequencies focusing on the effect of chirality.
Throughout the paper we consider the systems with the finite carrier density, i.e., either electron or hole doped (or gated) systems at zero temperatures.

We find many intriguing and unexpected features of the plasmon modes and their Landau damping in gapless semimetals. 
For chiral systems with the isotropic linear and quadratic band dispersion ($N=1, 2$ in $E\propto k^N$)
the long-wavelength plasmons 
lie outside the Landau damping regions due to the depolarization effects arising from the interband transition and therefore they do not decay by producing electron-hole pairs. 
For the systems with higher-order dispersions ($N \ge 3$), the long-wavelength 
plasma frequencies divided by Fermi energy (i.e., $\hbar\omega_{\rm p}/E_{\rm F}$) increase as the density decreases and enter the interband single particle excitation (SPE) region at the critical carrier density, where the dispersion shows a discrete energy jump.
For the multi-Weyl systems with an anisotropic dispersion (i.e., with a linear band dispersion along 
$z$ direction and non-linear in $x$-$y$ directions), one interesting salient feature in the calculated plasma frequency is the density dependence of the plasmon. We find that the density dependence of the long-wavelength plasma frequency along the direction of non-linear dispersion behaves like that of the isotropic linear band model ($N=1$), while along the direction of linear dispersion it behaves like that of the isotropic non-linear model ($N \ge 2$). We find that both the plasmons remain undamped over a broad range of density and interaction strength due to the chirality induced red-shift of plasmon modes.
We believe that our predictions may be easily observable via inelastic light-scattering spectroscopy\cite{Olego1982,Pinczuk1986,Eriksson1999} or inelastic electron-scattering spectroscopy \cite{Langer2010,Liu2008,Shin2011,Eberlein2008,Nagashima1992}.

The rest of the paper is organized as follows. In Sec. II, we consider semimetals with arbitrary isotropic band dispersion and present analytical and numerical results for long-wavelength plasma frequencies at zero temperature. In Sec. III, we consider multi-Weyl semimetals with an anisotropic dispersion and present results for zero-temperature plasma frequencies. In Sec. IV, we summarize and conclude our results.

\section{The Isotropic Model}

Before we calculate the collective modes in gapless multi-Weyl semimetals with anisotropic energy band dispersions, we first consider the gapless semimetals with the isotropic band dispersions.
For the isotropic model we consider the following Hamiltonian that describes a chiral gapless system with the symmetric energy dispersion, 
\begin{equation}\label{isotropic_model}
H_N^{\rm ch}=E_0\left(\frac{|{\bm k}|}{k_0}\right)^N \hat{\bm k}\cdot \bm{\sigma},
\end{equation}
where ${\bm k}$ is the wave vector, $\hat{\bm k}={\bm k}/|{\bm k}|$, $\bm{\sigma}$ are Pauli matrices acting in the space of two bands involved at the Weyl point, and $E_0$ and $k_0$ are material dependent parameters, which have units of energy and momentum, respectively. The corresponding energy dispersions are given by $E_{\bm{k},\pm}=\pm E_0(|{\bm k}|/k_0)^N$ and the eigenfunctions corresponding to the $\pm$ energies are 
\begin{eqnarray}
\left|+\right\rangle & = &(\cos{\theta\over 2},e^{i\phi}\sin{\theta\over 2}), \\
% for the conduction band and 
\left|-\right\rangle & = &(-\sin{\theta\over 2},e^{i\phi}\cos{\theta\over 2}),
\end{eqnarray}
where $\theta=\tan^{-1}\left(\sqrt{k_x^2+k_y^2}/k_z\right)$ and $\phi=\tan^{-1}\left(k_y/k_x\right)$. 
The $\left|+\right\rangle$ state with positive energy represents the conduction band, and the $\left|-\right\rangle$ state with negative energy represents the valence band.
Note that the Hamiltonian in Eq.~(\ref{isotropic_model}) with $N=1$ corresponds to the Hamiltonian of Weyl semimetals. 
To understand the effects of chirality on the plasmon properties beyond the energy dispersion of the system, we also consider the following form of a non-chiral gapless Hamiltonian:
\begin{equation}\label{non_chiral_isotropic_model}
H_N^{\rm nch}=E_0\left(\frac{|{\bm k}|}{k_0}\right)^N \sigma_z.
\end{equation}
The non-chiral Hamiltonian we introduce in Eq.~(\ref{non_chiral_isotropic_model}) has exactly the same energy dispersion as the chiral system in Eq.~(\ref{isotropic_model}). Unlike the chiral model, however, its eigenstates are completely independent of each other so that the chiral nature of wave functions is absent. It will be shown later that the interband transitions associated with chirality are largely responsible for differences between plasmons in the presence and absence of chirality.

Plasmons are defined as longitudinal in-phase oscillation of all the carriers driven by the self-consistent electric field generated by the local variation in charge density. To find the full plasmon dispersion at finite wave vectors we need the quantum mechanical many-body theory for the collective motion of all carriers\cite{Mahan2000}. Within the random phase approximation (RPA), the plasmon dispersion is obtained by finding zeros of the dynamical dielectric function, which is expressed as \cite{Mahan2000,Ando1982,Hwang2007,DasSarma2009, Zhou2015, Hofmann2015-1}
\begin{equation}\label{dielectric function}
\varepsilon^{\rm RPA}(q,\omega)=1-v_{\rm C}(q)\Pi(q,\omega),
\end{equation}
where $v_{\rm C}(q)={4\pi e^2 \over \kappa q^2}$ is the Coulomb interaction and $\kappa$ is a background dielectric constant. The non-interacting polarizability $\Pi(q,\omega)$ is given by
\begin{equation}\label{polarization function}
\Pi(q,\omega)=g\sum_{s,s'}\int\frac{d^3k}{(2\pi)^3}\frac{f_{\bm k,s}-f_{\bm k+\bm q,s'}}{\hbar\omega+\Delta_{{\bm k},{\bm k+\bm q}}^{ss'}+i0^{+}} F_{{\bm k},{\bm k+\bm q}}^{ss'},
\end{equation}
where $f_{\bm k,s}=\left[1+\exp\left({E_{\bm k,s}-\mu\over k_{\rm B}T}\right)\right]^{-1}$ is the Fermi distribution function for the band $s=\pm 1$, $\mu$ is the chemical potential, $\Delta_{{\bm k},{\bm k+\bm q}}^{ss'}=E_{\bm{k},s}-E_{\bm{k}+\bm{q},s'}$, $F_{{\bm k},{\bm k+\bm q}}^{ss'}=\frac{1}{2}(1+ss'\cos{\theta_{\bm k, \bm k+\bm q}})$ is the overlap factor, $\theta_{\bm k, \bm k+\bm q}$ is the angle between $\bm k$ and $\bm k+\bm q$, and $g$ denotes the number of degenerate Weyl nodes of the system.
%spin/valley degeneracy 
Since we focus on the long-wavelength plasmons in this work, we neglect transitions between different nodes located at different momenta. Within this assumption, contributions from other Weyl nodes can be taken into account by multiplying the degeneracy factor $g$. 
%Note that since only long-wavelength plasmons are considered in this work, transitions between nodes located at different momenta are negligible.  Thus, contributions from multiple nodes can be taken into account by multiplying the degeneracy factor $g$.
It is important to notice that in the non-chiral model described by Eq.~(\ref{non_chiral_isotropic_model}) and later by Eq.~(\ref{non_chiral_anisotropic_model}), the overlap factor becomes the Kronecker $\delta$, i.e., $F_{{\bm k},{\bm k+\bm q}}^{ss'}=\delta_{ss'}$, which excludes the possibility of any interband transitions between the conduction and valence bands. In this sense, the non-chiral model considered here is a two-band model without interband transitions, which is effectively a single-band model with the same energy dispersion.

\begin{figure}[htb]
\centering
\includegraphics[width=1\linewidth]{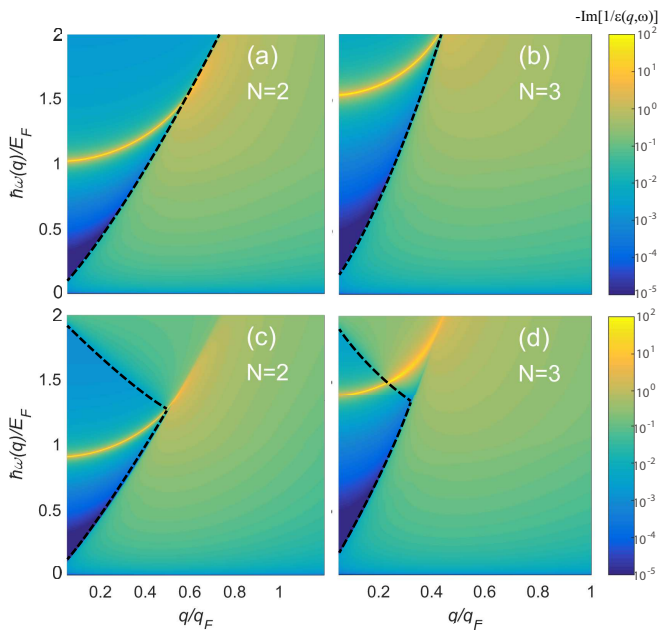}
\caption{
The density plots of calculated energy-loss functions $-\mathrm{Im}[1/\varepsilon(q,\omega)]$ in ($q$,$\omega$) space. (a) and (b) show the loss functions in the absence of chirality for (a) $N=2$ (parabolic dispersion) and (b) $N=3$ (cubic dispersion), whereas (c) and (d) show the energy-loss functions in the presence of chirality for (c) $N=2$ and (d) $N=3$. The dashed lines represent the boundaries of the intraband and interband electron-hole continua.
Note that in the presence of chirality, the plasmon energies are shifted downward for all wave vectors compared with those in the absence of chirality. Here we use the coupling constant $g\alpha=2.4$, and the Fermi energy $E_{\rm F}/E_0=1$ for $N=2$ and $E_{\rm F}/E_0=0.5$ for $N=3$, respectively.}
\label{fig:1}	
\end{figure}

We first consider the loss function of 3D gapless semimetals, which is calculated from the dielectric function (i.e., $-\mathrm{Im}[1/\varepsilon(q,\omega)]$) and can be directly measured in experiments such as inelastic electron spectroscopy. 
The loss function is related to the dynamical structure factor $S(q,\omega)$ by $S(q,\omega) \propto -\mathrm{Im}[1/\varepsilon(q,\omega)]$, which gives a direct measure of the spectral strength of the various elementary excitations. 
%Thus, our calculated loss function can be measured in experiments such as inelastic electron and Raman-scattering spectroscopies. 
Figure \ref{fig:1} shows the density plots of calculated energy-loss functions in ($q,\omega$) space for $N=2$ (parabolic dispersion) and $N=3$ (cubic dispersion) in the absence (top panels) and presence (bottom panels) of chirality. 
Plasmon dispersions are given by sharp peaks of the energy-loss function, which correspond to the poles of the dielectric function. 
When both Re[$\epsilon$] and Im[$\epsilon$] become zero (i.e., $\epsilon(q,\omega)= 0$, which defines the plasmon mode), the imaginary part of the inverse dielectric function becomes the Dirac $\delta$-function, i.e., $-\mathrm{Im}[1/\varepsilon(q,\omega)]=W(q) \delta(\omega-\omega_{\rm p}(q))$ with the strength
\begin{equation}
W(q) = \pi \left \{ \partial {\rm Re}[\epsilon(q,\omega)]/\partial \omega|_{\omega=\omega_{\rm p}(q)} \right \}^{-1},
\end{equation} 
where $\omega_{\rm p}(q)$ is the plasma frequency at a given wave vector $q$.
Thus, an undamped plasmon shows up as a well-defined $\delta$-function peak in the loss function
as indicated by sharp yellow solid lines in Fig.~\ref{fig:1}. The undamped plasmon mode in general carries most of spectral weights and should be observable in experiments (i.e., it is expected that the mode does not decay by electron-hole pairs).
The dotted lines in Fig.~\ref{fig:1} represent the boundaries of the intraband and interband electron-hole single particle excitation (SPE) continua. The electron-hole SPE continua show up as weak broad incoherent structure and carries small spectral weight. 
When the plasmon mode enters the SPE continuum at the critical wave vector $q_{\rm c}$, the dielectric function $\epsilon(q>q_{\rm c}, \omega)$ has a finite imaginary part and the plasmon mode becomes damped via Landau damping. The plasmon mode inside the Landau damping region decays by emitting intraband or interband electron-hole pairs, which is now allowed by energy-momentum conservation. The broadened peaks inside the SPE regions in Fig.~\ref{fig:1} indicate the damped plasmons. 
The plasmon energy scaled by the Fermi energy ($\hbar\omega_{\rm p}/E_{\rm F}$) is strongly dependent on the band structure. 
One interesting result is the chirality dependence of the plasmon energy, and it is important to note that in the presence of chirality, plasma frequencies are red-shifted. This is due to the depolarization effect on the plasmon modes, arising from interband transitions.

\begin{figure}[htb]
\centering
\includegraphics[width=1\linewidth]{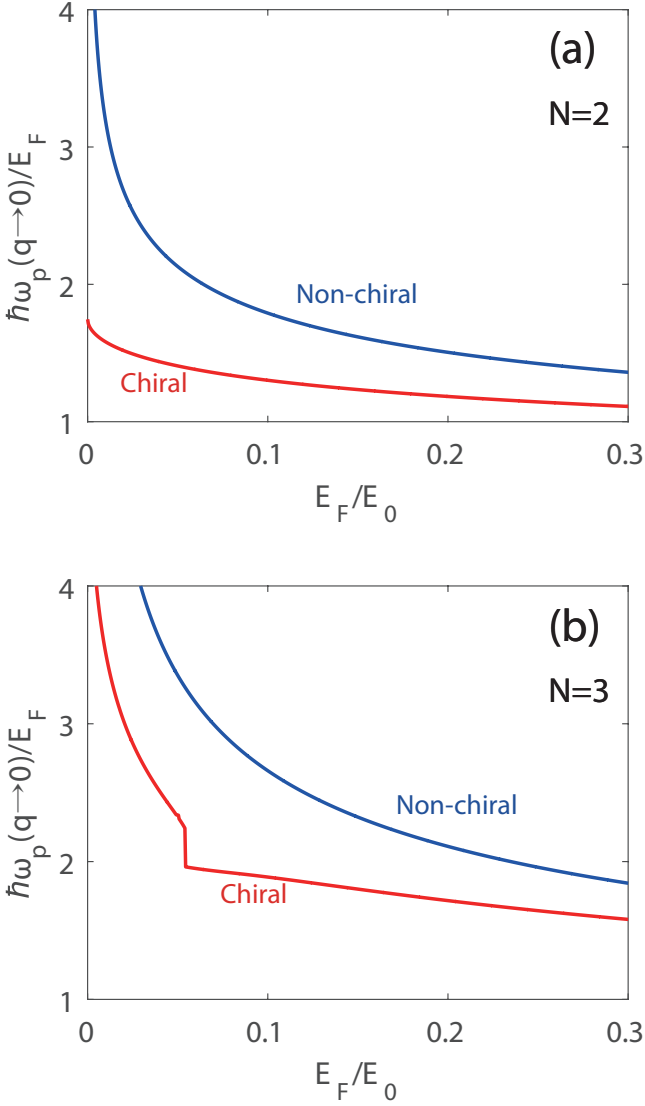}		
\caption{
Density dependence of the long-wavelength plasma frequencies for (a) $N=2$ and (b) $N=3$. Red (Blue) lines correspond to the results in the presence (absence) of chirality. 
Note that for $N=2$, $\hbar \omega_{\rm p}/E_{\rm F}$ diverges in the $E_{\rm F}\rightarrow0$ limit in the absence of chirality, whereas it converges to a finite value in the presence of chirality. For $N=3$, the chiral plasmon dispersion shows a discrete jump at $\hbar\omega_{\rm p}=2E_{\rm F}$, where plasmons start decaying into electron-hole pair excitations. Here we use the coupling constant $g\alpha=2.4$.}
\label{fig:2}	
\end{figure}

Figure \ref{fig:2} shows the long-wavelength plasma frequencies for $N=2$ and $N=3$ as a function of Fermi energy in the presence (red) and absence (blue) of chirality. In this figure the boundary of interband SPE at $q=0$ is $2E_{\rm F}$. If the plasma frequency is larger (smaller) than $2E_{\rm F}$, the plasmon decays (does not decay)  by Landau damping. 
For both $N=2$ and $N=3$, $\hbar\omega_{\rm p}/E_{\rm F}$ in the absence of chirality increases monotonically with decreasing density and diverges in the low density limit ($E_{\rm F}\rightarrow0$). 
In the presence of chirality, the plasma frequencies are always smaller than those without chirality, and the energy difference between the two plasmons grows as the carrier density decreases. For $N=2$, as the density goes to zero, the plasma frequency approaches a finite value less than $2E_{\rm F}$, i.e., $\hbar\omega_{\rm p}\simeq1.7 E_{\rm F}$, 
indicating that the long-wavelength plasmon does not enter the interband electron-hole continuum %($\hbar\omega_{\rm p}>2E_{\rm F}$) 
for the whole range of densities and thus the plasmon is not damped through the Landau damping. For $N=3$, the plasma frequency  as a function of Fermi energy exhibits a discrete jump at a critical value of $E_{\rm F} = \hbar\omega_{\rm p}/2$, and below the critical value the plasmon energy becomes bigger than $2E_{\rm F}$ and enters into the interband electron-hole continuum decaying via Landau damping.
Note that interband transitions are more suppressed at high densities 
due to Pauli-blocking, which narrows the energy range over which interband transitions can occur. 
Thus, the effect of interband transition on plasmons becomes weaker at high densities, resulting in small depolarization shift in plasmon energy. 
For this reason, the signature of chirality in the plasmon dispersions is significant at low densities.

To get further insight into the results shown in Fig.~\ref{fig:1} , here we present the calculated analytic expressions for the leading-order long-wavelength plasma frequencies (for details, see Appendix A),
\begin{equation} \label{nonchiral plasma frequency}
\hbar\omega^{\rm nch}_{\rm p}=E_0\sqrt{\frac{2Ng\alpha}{3\pi}} \left(\frac{E_{\rm F}}{E_0}\right)^{\frac{1+N}{2N}}
\end{equation}
in the absence of chirality and 
\begin{equation} \label{chiral plasma frequency}
\hbar\omega^\mathrm{ch}_{\rm p}
=
\hbar \omega^\mathrm{nch}_{\rm p}\left[1+\frac{g\alpha}{3\pi(N-1)}\left(\frac{E_{\rm F}}{E_0}\right)^{\frac{1}{N}-1}\right]^{-\frac{1}{2}}
\end{equation}
in the presence of chirality, where $\alpha={e^2 k_0\over \kappa E_0}$ is the coupling constant characterizing the interaction strength. 
The above results are calculated in $E_{\rm F}/E_0\ll1$ limit for $N\geq 2$.
Note that for $N=1$ Weyl semimetals a linear band dispersion leads to cut-off dependent long-wavelength plasma frequencies\cite{Hofmann2015-2, Throckmorton2015}.
It is easy to see from Eq.~(\ref{chiral plasma frequency}) that the plasma frequency with chirality is red-shifted with respect to that without chirality, showing different density dependence at low densities. This difference originates from the interband transition contribution, 
%which appears in the second term of the polarizability with the opposite sign of the first term arising from the intraband contribution 
which appears in the additional term of the polarizability with the opposite sign of the intraband contribution 
(see Eq.~(A8) in Appendix A). This indicates that interband transitions and associated chiral nature of wavefunctions contribute to the depolarization of the screening and are responsible for the red-shift of the plasma frequencies. 

It is interesting that the density dependence of the long wavelength plasma frequencies shown in Eq.~(\ref{nonchiral plasma frequency})
can be obtained from
the classical plasma frequency \cite{Jackson1999}
\begin{equation}\label{classical plasmon}
\omega^{\rm cl}_{\rm p}=\sqrt{\frac{4\pi n e^2}{\kappa m}},
\end{equation}
%where $n$ is the charge carrier density, $m$ is the effective mass of the charge carrier, and $\kappa$ is a background dielectric constant of the materials. 
where $n$ is the charge carrier density and $m$ is the effective mass of the charge carrier. 
By using the momentum relation $m v_{\rm F} = \hbar k_{\rm F}$ ($v_{\rm F}$ and $k_{\rm F}$ are the Fermi velocity and the Fermi wave-vector, respectively) and the energy dispersion relation $E_{\bm{k},\pm}=\pm E_0(|{\bm k}|/k_0)^N$
for arbitrary band dispersion, 
the density dependence of the classical plasma frequency  can be calculated as
\begin{equation}\label{classical_plasmon_density_dependence}
\hbar\omega^{\rm cl}_{\rm p}\sim E_{\rm F}^\frac{1+N}{2N},
\end{equation}
which agrees with the full RPA result in Eq.~(\ref{nonchiral plasma frequency}) for the non-chiral case.
Thus, for $N=1$ we have $\hbar\omega_{\rm p}^{\rm cl} \sim E_{\rm F}$, i.e., the plasma frequency scaled by the Fermi energy is independent of $E_{\rm F}$\cite{DasSarma2009}.
For $N\ge 2$, $\hbar\omega_{\rm p}^{\rm cl}/E_{\rm F} \sim E_{\rm F}^{-\alpha}$ with $\alpha = (N-1)/2N$, i.e., the scaled plasma frequency increases as the density (or Fermi energy) decreases and diverge as $E_{\rm F} \rightarrow 0$. Note that the interband transition red-shifts the classical plasma frequencies in the presence of chirality.

\section{The Anisotropic Model}

In the previous section we have discussed the plasmon properties of isotropic gapless semimetals.
In this section we explore, within the RPA,  the plasmon properties of  the anisotropic multi-Weyl system whose dispersion is non-linear in the in-plane directions, but linear in the  out-of-plane direction.

We consider the following Hamiltonian that describes a multi-Weyl node of order $J$,
\begin{equation}\label{anisotropic_model}
%H_J^{\rm ch}(\bm{k})= c_\parallel ( k_-^J\sigma_+ + k_+^J\sigma_-) + c_z k_z\sigma_z,
%H_J^{\rm ch}(\bm{k})= E_0[(\tilde{k}_-^J\sigma_+ + \tilde{k}_+^J\sigma_-) + \tilde{k_z}\sigma_z],
H_J^{\rm ch}(\bm{k})= E_0\left[\left({k_{-}\over k_0}\right)^J\sigma_{+} + \left({k_{+}\over k_0}\right)^J\sigma_{-} + {k_z\over k_0}\sigma_z\right],
\end{equation}
where $\sigma_\pm=\frac{1}{2}\left(\sigma_x\pm i\sigma_y\right)$, $k_{\pm}=k_x\pm i k_y$, and $k_0$ and $E_0$ are material dependent parameters.
Throughout this section the momentum and the energy are normalized by $k_0$ and $E_0$, respectively. With these normalized quantities, we can write the energy dispersion as $E_{\bm{k},\pm}=\pm E_{\bm{k}}$, where $E_{\bm{k}}=\sqrt{k_{\parallel}^{2J}+k_z^2}$, $k_{\parallel}=\sqrt{k_x^2+k_y^2}$ is the in-plane momentum corresponding to non-linear dispersion ($E\sim k_\parallel^J$) in the $x-y$ plane and $k_z$ is the out-of-plane momentum along the $z$ direction where the dispersion is linear ($E\sim k_z$).
The overlap factor $F^{ss'}_{\bm k,\bm k'}$ is given by  
\begin{equation}
F^{ss'}_{\bm k,\bm k'}={1\over 2}\left[1+ss'(\cos{\theta}\cos{\theta'}+\sin{\theta}\sin{\theta'}\cos J(\phi-\phi'))\right]
\end{equation}
where $\bm k =(k_x, k_y, k_z)$ and $\bm k' =(k_x', k_y', k_z')$ are related to $(r,\theta,\phi)$ and $(r',\theta',\phi')$ through the coordinate transformation, respectively (see Eq.~(B1) in Appendix B).
For comparison, we also introduce an anisotropic non-chiral model with the same energy dispersion:
\begin{equation}\label{non_chiral_anisotropic_model}
H_J^{\rm nch}(\bm{k})= E_{\bm{k}}\sigma_z.
\end{equation}

\begin{figure}[htb]	
\centering
\includegraphics[width=1\linewidth]{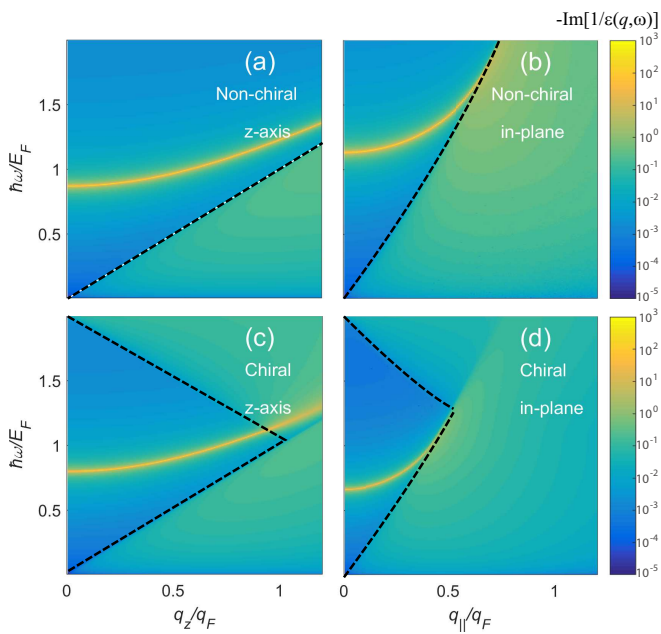}
\caption{
Calculated plasmon dispersions (solid yellow lines) for the anisotropic model with $J=2$ in the absence [top panels, (a) and (b)] and presence [bottom panels, (c) and (d)] of chirality along the $z$-direction [left panels, (a) and (c)] and the in-plane direction [right panels, (b) and (d)]. Note that plasmon dispersions along the $z$-direction behave similarly as those for the $N=1$ isotropic model and plasmon dispersions along the in-plane direction behave as those for the $N=2$ isotropic model. Plasmon frequencies along both directions are red-shifted in the presence of chirality, as in the isotropic model. Here we use the Fermi energy $E_{\rm F}/E_0=1$ and the coupling constant $g\alpha=2.4$.
}
\label{fig:3}
\end{figure}

To investigate the plasmon dispersions and  their damping we calculate the energy loss function of the system. Figure \ref{fig:3} shows the density plots of the calculated energy loss functions for $J=2$ in the absence (top panels) and presence (bottom panels) of chirality. Figure \ref{fig:3} (a) and (c) show the density plots of the energy loss function in ($q_z,\omega$) space for a fixed in-plane wave vector ($q_\parallel=0$), and
Fig.~\ref{fig:3} (b) and (d) show the density plots of the energy loss function in ($q_\parallel,\omega$) space for a fixed out of plane wave vector ($q_z=0$). The yellow lines with the largest spectral weight in the energy loss correspond to the plasmon dispersions of the system.
The two most important salient features of the results shown in Fig.~\ref{fig:3} are following. First, the plasmon dispersions for both in-plane and out-of-plane directions are red-shifted in the presence of chirality, which is consistent with the isotropic results. Second, the plasmon dispersion along the out-of-plane direction (where the band dispersion is linear) shows a similar behavior as that in the $N=1$ isotropic system.
This is because the longitudinal plasmon oscillations propagating along the out-of-plane direction arise from the collective carriers with linear band dispersion and thus their effective motion is essentially identical to those in gapless semimetals with linear dispersion ($N=1$). The same argument is also applied to the in-plane plasmon modes which propagate along the in-plane direction, where the carriers have the parabolic dispersion and therefore the plasmon dispersion behaves as that in the $N=2$ isotropic system.

\begin{figure}[htb]
\centering
\includegraphics[width=1\linewidth]{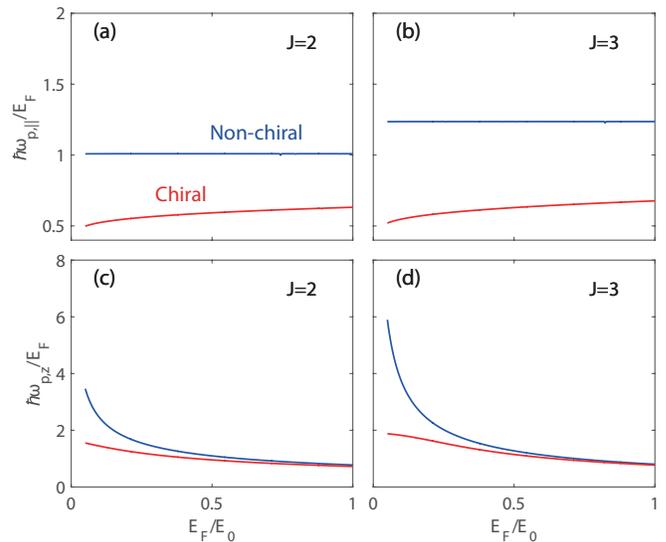}	
\caption{
Density dependence of the long-wavelength plasma frequencies in the presence (red lines) and absence (blue lines) of chirality for $J=2$ [left panels, (a) and (c)] and $J=3$ [right panels, (b) and (d)].
Along the in-plane momentum direction [top panels, (a) and (b)], $\hbar\omega_{\rm p}/E_{\rm F}$ for both $J=2$ and $J=3$ behaves as that in Weyl semimetals. Along the $z$ direction [bottom panels, (c) and (d)], for both $J=2$ and $J=3$ $\hbar\omega_{\rm p}/E_{\rm F}$ converges to a finite value in the $E_{\rm F}\rightarrow 0$ limit. Here we use the coupling constant $g\alpha=2.4$.
}
\label{fig:4}
\end{figure}

In contrast to the plasmon dispersions, however, the density dependence of the long-wavelength plasmons shows a non-trivial relation between the isotropic and anisotropic models.
Figure \ref{fig:4} presents calculated long-wavelength plasma frequencies as a function of Fermi energy for $J=2$ and $J=3$ in the presence (red) and absence (blue) of chirality. 
The in-plane plasma frequencies $\hbar\omega_{\rm p}/E_{\rm F}$ [(a) for $J=2$ and (b) for $J=3$],  show the same behavior as the $N=1$ isotropic system, i.e., the normalized frequencies are weakly dependent on the carrier density over a wide range both with and without chirality. Note that this behavior is similar to that in gapless semimetals with linear dispersion. 
For the $z$-direction [(c) for $J=2$ and (d) for $J=3$], $\hbar\omega_{\rm p}/E_{\rm F}$ increases with decreasing Fermi energy in the absence of chirality, and diverges in the $E_{\rm F}\rightarrow0$ limit for both $J=2$ and $J=3$. In the presence of chirality, for $J=2$ and $J=3$ it converges to a finite value in the $E_{\rm F}\rightarrow0$ limit. Note that these results are qualitatively similar to the chiral signature of $N=2$ and $N=3$ isotropic systems, respectively, as discussed in Sec. II, though here the energy dispersion along the $z$-direction is linear. 
It is important to notice that this result is against our expectations because the density dependence of the anisotropic plasmons along a direction with a specific energy dispersion does not match with the corresponding  plasmons in the isotropic model with the same energy dispersion. 
It should also be noted that for the chiral case anisotropic plasmons along both the in-plane and out-of-plane directions are significantly red-shifted so that they are outside the electron-hole continua for a broad range of density and coupling strength. For $J=3$ plasmons, this is in sharp contrast to the counterpart of the isotropic case ($N=3$), where plasmons exhibit a discrete jump merging into the electron-hole continua at a critical Fermi energy.

Here we analyze the previous numerical results in Fig.~\ref{fig:4} with analytical expressions. Similarly as the isotropic case, we can also derive long-wavelength plasma frequencies from the linear response theory in the many-body approach. In the absence of chirality, we obtain plasma frequencies for the anisotropic model to be 
\begin{equation}\label{anisotropic_plasmon_xy}
\omega_{{\rm p},\parallel}= \sqrt{ \frac{2 g\alpha J }{3 \pi}} E_{\rm F}
\end{equation}
along the $k_\parallel$ momentum direction and
\begin{equation} \label{anisotropic_plasmon_z}
\omega_{\rm p}=\sqrt{\frac{g\alpha \Gamma \left(1+\frac{1}{J}\right)}{2\sqrt{\pi}\Gamma \left(\frac{3}{2}+\frac{1}{J}\right)}} E_{\rm F}^{1/J}
\end{equation}
along the $k_z$ momentum direction (For details, see Appendix B).
The difference in density dependence between the two plasma frequencies can also be understood by the classical derivation of plasma frequency, as in the isotropic case.  Since the density of states of the anisotropic model in Eq.~(\ref{anisotropic_model}) is calculated to be $D(E)\sim E^{2/J}$, the total carrier density $n$ becomes $n\sim E_{\rm F}^\frac{2+J}{J}$. 
Since the energy dispersion is anisotropic, it is expected that the effective mass is also anisotropic.
%To obtain the effective mass, we need to consider the energy dispersion along the propagating direction, because the longitudinal plasmon wave oscillates in the direction of propagation. 
Thus, electrons would effectively behave as if they have the energy dispersion $E\sim k_\parallel^J$ with $m\sim E_{\rm F}^{{2-J\over J}}$ along the in-plane $k_\parallel$ momentum direction, and $E\sim k_z$ with $m\sim E_{\rm F}$ for the out-of-plane $k_z$ momentum direction. 
Putting $n$ and $m$ into Eq.~(\ref{classical plasmon}), we can obtain the Fermi energy dependence of the plasma frequency in classical limit:
\begin{equation}
\hbar\omega_{{\rm p},\parallel}\sim E_{\rm F}
\end{equation}
along the $k_\parallel$ momentum direction and
\begin{equation}
\hbar\omega_{{\rm p},z}\sim E_{\rm F}^{1/J}
\end{equation}
along the $k_z$ momentum direction.
The density dependence of plasmons obtained from both classical and quantum mechanical approaches are consistent with the numerical results in Fig.~\ref{fig:4}.
Note that as in the case of the isotropic system, the density dependence of classical plasma frequencies is in good agreement with that of the non-chiral plasmon.

\section{Summary and Conclusion}

In this paper we investigate theoretically electronic collective modes of  3D chiral gapless electron-hole systems % with arbitrary energy band dispersion in three dimensions 
and find the wave vector dependent plasmon dispersion. We have calculated long-wavelength plasma frequencies and their density dependence both in semimetals with an isotropic band dispersion and in multi-Weyl semimetals with an anisotropic band dispersion.
% focusing on the effects of interband transitions.
%We consider both chiral and non-chiral cases to compare the effect of chirality on plasmon properties. 
We find that the interband transition associated with chirality leads to the depolarization shift of plasma frequencies irrespective of band dispersion.

For the isotropic parabolic dispersion ($N=2$), the depolarization shift of the long-wavelength plasmons arises from the interband electron-hole transition and the plasmons lie outside the interband electron-hole continuum (i.e., $\hbar \omega_{\rm p} < 2E_{\rm F}$). Thus, the plasmons do not decay via Landau damping. 
For the cubic dispersion ($N=3$), we find that the plasma frequency ($\hbar \omega_{\rm p}/E_{\rm F}$) increases
as the density decreases and enters the interband single particle excitation region at a critical carrier density showing a discrete energy jump.

For the anisotropic dispersion, we have calculated the plasma frequencies in a system with a linear band dispersion along one specific direction and non-linear dispersion in the other remaining directions. We find that the density dependence of the long-wavelength plasma frequency along the direction of non-linear dispersion shows a similar behavior as that of the linear band isotropic model ($N=1$), in which $\hbar\omega_{\rm p}/E_{\rm F}$ depends weakly on the density over a wide range of Fermi energies. On the other hand, the density dependence of the long-wavelength plasma frequency along the direction of linear dispersion shows a similar behavior as that of the isotropic model with the non-linear dispersion. %, exhibiting the chiral characteristics arising from interband transitions.
%At long wavelengths $q\rightarrow 0$, the density dependence of the plasma frequencies of anisotropic systems is different from the isotropic gapless electron-hole system. 
The long-wavelength plasmons along all the directions are undamped due to the chirality induced red-shift of plasma frequencies.

Our predicted plasmon properties clearly distinguish gapless semimetals from the extensively studied usual parabolic 3D electron systems. We believe that our predictions can be tested in doped gapless semimetals using inelastic light scattering \cite{Eriksson1999, Pinczuk1986, Olego1982} and electron scattering \cite{Langer2010, Liu2008, Shin2011, Eberlein2008, Nagashima1992} spectroscopies.

%%%%%%%%%%%%%%%%%%%%%%%%%%%%%%%%%%%%%%%%%%%%%%%%%%%%%%%%%%%%%%%%%%%%%%%%%%%%%%%%%%%%%%%%%%%%%%%%%%%%

\acknowledgments 
This research was supported by Basic Science Research Program through the National Research Foundation of Korea (NRF) funded by the Ministry of Education under Grant No. 2015R1D1A1A01058071 (HM) and 2014R1A2A2A01006776 (EHH).

\vspace*{0.5cm}

\appendix

\section{Derivation of the long-wavelength polarization function for the isotropic model}
In this section, we show the derivation of the analytical expressions for the zero-temperature polarization function and plasma frequencies in the $q\rightarrow0$ limit. The polarization function for the isotropic model can be written as
\begin{equation}
\Pi(q,\omega)=g\sum_{s,s'=\pm}\Pi_{ss'}(q,\omega),
\end{equation}
where
\begin{equation}
\Pi_{ss'}(q,\omega)
=\int\frac{d^3k}{(2\pi)^3}\frac{f_{\bm k,s}-f_{\bm k+\bm q,s'}}{\hbar\omega+\Delta_{{\bm k},{\bm k+\bm q}}^{ss'}+i0^{+}} F_{{\bm k},{\bm k+\bm q}}^{ss'}.
\end{equation}

Note that at zero temperature the Fermi distribution function is given by $f_{\bm k,s}=\Theta\left(E_{\rm F}-E_{\bm{k},s}\right)$, where $\Theta(x)$ is the step function [$\Theta(x)=0$ for $x<0$ and $\Theta(x)=1$ for $x\geq0$].
We can break the polarization function into two parts due to intraband transitions and interband transitions, respectively:
$\Pi(q,\omega)=
\Pi_{\mathrm{intra}}(q,\omega)+
\Pi_{\mathrm{inter}}(q,\omega)
$, where 
$\Pi_{\mathrm{intra}}(q,\omega)=\Pi_{++}(q,\omega)+\Pi_{--}(q,\omega)$ and 
$\Pi_{\mathrm{inter}}(q,\omega)=\Pi_{+-}(q,\omega)+\Pi_{-+}(q,\omega)$. 
For convenience, we use the change of variable $\bm{k}\rightarrow-\bm{k-q}$ and introduce dimensionless quantities $x=k/k_{\rm F}, y=q/k_{\rm F}, z=\hbar\omega /E_{\rm F}$ and $\Pi(q,\omega)=D(E_{\rm F})\widetilde{\Pi}(y,z)$, where $D(E_{\rm F})$ is the density of states at the Fermi energy given by $D(E_{\rm F})=\frac{gk_0^3}{2\pi^2NE_0}(E_{\rm F}/E_0)^{{(3-N)}/{N}}$.

Then we can rewrite $\widetilde{\Pi}_{\mathrm{intra}}(q,\omega)$ and $\widetilde{\Pi}_{\mathrm{inter}}(q,\omega)$ as
\begin{widetext}
\begin{equation}\label{eq:intra}
\widetilde{\Pi}_{\mathrm{intra}}(y,z)
=\frac{N}{2}
\int_0^1 x^2dx
\int_0^\pi \sin\theta d\theta
\left(
\frac{1}{z-\Delta_{\mathrm{+}}}
-
\frac{1}{z+\Delta_{\mathrm{+}}}
\right)
F_{\mathrm{+}}
\end{equation}
and
\begin{equation}\label{eq:inter}
\widetilde{\Pi}_{\mathrm{inter}}(y,z)
=\frac{N}{2}
\int_1^\infty x^2dx
\int_0^\pi \sin\theta d\theta
\left(
\frac{1}{z-\Delta_{\mathrm{-}}}
-
\frac{1}{z+\Delta_{\mathrm{-}}}
\right)
F_{\mathrm{-}}
\end{equation}
\end{widetext}
where
$\Delta_{\pm}=(x^2+y^2+2xy\cos{\theta})^\frac{N}{2}\mp x^N $. In the $q\rightarrow0$ limit $(z\pm\Delta_\pm)^{-1}\approx z^{-1}\left[1\mp \frac{\Delta_\pm}{z}\pm\left(\frac{\Delta_\pm}{z}\right)^2+\cdots\right]$ and $\Delta_\pm\approx (x^N\mp x^N)+N  \cos \theta  x^{N-1} y+\frac{1}{2} N \left\{1+(N-2)\cos^2\theta\right\}x^{N-2} y^2+\cdots$. In the presence of chirality the overlap factor is given by $F_{\pm}=\frac{1\pm \cos{\theta'}}{2}\approx  \frac{1\pm1}{2} \mp \frac{\sin^2\theta}{4 x^2}y^2+\cdots$, where $\cos\theta'=\frac{x+y\cos\theta}{\sqrt{x^2+y^2+2xy\cos\theta}}\approx 1-\frac{\sin\theta^2}{2x^2}y^2$ whereas in the absence of chirality $F_{+}=1$ and $F_{-}=0$. 
By putting all these equations into Eq.~(\ref{eq:intra}) and expanding up to the second order in $y$, we can obtain the long-wavelength polarization function. For the intraband part $\widetilde{\Pi}_{\mathrm{intra}}(y,z)$, the polarization functions for the chiral and non-chiral cases are the same up to the second order, given by
\begin{equation}
\begin{split}
\widetilde{\Pi}_{\mathrm{intra}}(y,z) \approx\frac{N}{2} \int_0^1 x^2dx \int_0^\pi \sin\theta d\theta \frac{2\Delta_{\mathrm{+}}}{z^2} F_{\mathrm{+}}
=\frac{N^2 y^2}{3 z^2}+O(y^3).
\end{split}
\end{equation}
For the interband part $\widetilde{\Pi}_{\mathrm{inter}}(y,z)$,  note that there is no interband contribution for the non-chiral case due to the absence of interband transitions ($F_-=0$). For the chiral case, $F_-\approx (\sin^2\theta/4 x^2)y^2$ and when we expand the integrand only up to the second order, we can approximate $\Delta_-=2x^N$, yielding
\begin{widetext}
\begin{equation}
\begin{split}
\widetilde{\Pi}_{\mathrm{inter}}y,z)
&=
\frac{N}{2}
\int_1^\infty x^2dx
\int_0^\pi \sin\theta d\theta
\left(
\frac{4x^{N}}{z^2-(2x^{N})^2}
\right)
\frac{\sin^2\theta}{4 x^2}y^2 + O(y^3)
=
\frac{2Ny^2}{3}
\int_1^\infty dx
\frac{x^{N}}{z^2-4x^{2N}} + O(y^3)\\
=&
-\frac{N \, _2F_1\left(1,\frac{N-1}{2 N};\frac{3N-1}{2N};\frac{z^2}{4}\right)}{6 (N-1)}y^2+O(y^3) \text{  for $z<2$},
\end{split}
\end{equation}
\end{widetext}
%where $_2F_1\left(1,\frac{N-1}{2 N};\frac{3}{2}-\frac{1}{2 N};\frac{z^2}{4}\right)=\sum_{m=0}^{\infty}\frac{ (z/2)^{2m}(N-1)}{[(2m+1)N-1]}$.
where $_2F_1\left(a,b;c;z\right)=\frac{\Gamma(c)}{\Gamma(b)\Gamma(c-b)}\int^1_0 dt\frac{t^{b-1}(1-t)^{c-b-1}}{(1-tz)^a}$ is the hypergeometric function, which can be obtained after substituting $t=x^{-2N}$.

Thus the full polarizability in the long wavelength limit ($q\rightarrow0$) is given by 
\begin{equation}\label{nonchiral polarization function}
\widetilde{\Pi}^{\rm nch}(y,z)=\frac{N^2y^2}{3z^2}
\end{equation}
for the non-chiral case whereas
%By adding up $\widetilde{\Pi}_{\mathrm{intra}}(y,z)$ and $\widetilde{\Pi}_{\mathrm{inter}}(y,z)$, we can obtain the full $q\rightarrow0$ polarizability,
\begin{equation} \label{chiral polarization function}
\widetilde{\Pi}^{\rm ch}(y,z)=\frac{N^2y^2}{3z^2}-\frac{N \, _2F_1\left(1,\frac{N-1}{2 N};\frac{3N-1}{2N};\frac{z^2}{4}\right)}{6 (N-1)}y^2 \text{  for $z<2$}
\end{equation}
for the chiral case. Note that the polarization function diverges when $N=1$ because a linear band dispersion leads to the divergence of the polarization function at infinite cut-off\cite{Hofmann2015}.
By putting Eq.~(\ref{nonchiral polarization function}) and Eq.~(\ref{chiral polarization function}) into the dielectric function in Eq. (5) in the main text, we arrive at the analytic expressions for the leading-order long-wavelength plasma frequencies given in Eq. (8) and Eq. (9) in the main text.

\section{Derivation of the long-wavelength polarization function in the anisotropic model}
In this section, we show the derivation of the polarization function for the anisotropic case. %, i.e., Eq.~(\ref{eq:inter}). 
Here we consider only the non-chiral case, where the overlap factor is given by $F_+=1$ and $F_-=0$. Note that due to the anisotropic band structure it is no longer convenient to normalize the polarization function with the density of states as in the isotropic case. Thus for the anisotropic case we normalize the polarization function by $\widetilde{\Pi}(\widetilde{\bm{q}},\widetilde{\omega})=\Pi(\bm{q},\omega)/\mathcal{C}$, where $\mathcal{C}=\frac{g}{(2\pi)^3}\frac{k_0^3}{E_0}$,  $\widetilde{\bm{q}}=\frac{\bm{q}}{k_0}$ and $\widetilde{\omega}=\frac{\hbar\omega}{E_0}$. In the following, for brevity, the momentum and energy without tilde are considered to be normalized by $k_0$ and $E_0$, as in the main text.
%For brevity, the momentum and the energy are normalized by $k_0$ and $E_0$, respectively, as in the main text.

We consider the following coordinate transformation:
\begin{equation}\label{coord}
\begin{split}
k_x&=(r \sin\theta)^\frac{1}{J}\cos\phi,\\
k_y&=(r \sin\theta)^\frac{1}{J}\sin\phi,\\
k_z&=r\cos\theta.
\end{split}
\end{equation}
Note that with this coordinate transformation the energy dispersion becomes linear: $E_{\pm}(r)=\pm r$. The Jacobian corresponding to this transformation is given by $\mathcal{J}(r,\theta)=\frac{1}{J}\left(r\sin\theta\right)^\frac{2-J}{J}r$.

For the non-chiral case, there is no interband contribution, thus we consider only the intraband contribution. With a similar procedure as in the previous section, we can write the polarization function for the in-plane momentum $q_{\parallel}$ along the $y$-axis and out-of-plane momentum $q_z$ along the $z$-axis as
\begin{widetext}
\begin{equation}\label{eq:aniintra}
\widetilde{\Pi}(q_{\parallel,z},\omega)
=
\int_0^{E_{\rm F}} dr
\int_0^\pi d\theta
\int_0^{2\pi} d\phi
\mathcal{J}(r,\theta)
\left(
\frac{1}{\omega-\Delta_{\parallel,z}} - \frac{1}{\omega+\Delta_{\parallel,z}}
\right)
\end{equation}
where 
\begin{equation}
\Delta_\parallel=E_{\bm{k} + \bm{q}_\parallel}-E_{\bm k}
=\sqrt{r ^2 \cos ^2\theta+\left[(r  \sin \theta)^{\frac{2}{J}}+2 q_\parallel \sin \phi (r  \sin \theta)^{\frac{1}{J}}+q_\parallel^2\right]^J}-r
\end{equation}
and
\begin{equation}
\Delta_z=E_{\bm{k} + \bm{q}_z}-E_{\bm k}
=\sqrt{r^2+q_z^2+2 r q_z \cos\theta}-r.
\end{equation}
\end{widetext}
In the $q\rightarrow0$ limit, $(\omega\pm\Delta_{\parallel,z})^{-1}\approx \omega^{-1}[1\mp\frac{\Delta_{\parallel,z}}{\omega} \pm \left(\frac{\Delta_{\parallel,z}}{\omega}\right)^2+\cdots]$,
$\Delta_z\approx q_z \cos\theta+ \frac{\sin ^2\theta }{2 r} q_z^2+\cdots $
and 
$\Delta_\parallel\approx A_1 q_{\parallel} + A_2 q_{\parallel}^2 + \cdots$
where
$A_1= J \sin \theta \sin \phi (r  \sin \theta)^{\frac{J-1}{J}}$
and
$A_2=\frac{1}{4} J \sin \theta (r  \sin \theta )^{\frac{J-2}{J}} \left[J\sin^2\phi(\cos2\theta+3)+2\cos2\phi\right]$.

By putting all these into Eq.~(\ref{eq:aniintra}), we get the following polarization functions:
\begin{widetext}
\begin{eqnarray}
\label{eq:q_parallel}
\widetilde{\Pi}(q_\parallel,\omega)
&\approx& \int_0^{E_{\rm F}} dr \int_0^\pi d\theta \int_0^{2\pi} d\phi \mathcal{J}(r,\theta) \frac{2\Delta_{\parallel}}{\omega^2}
\approx \frac{J\pi q_{\parallel}^2}{2\omega^2} \int_0^{E_{\rm F}} dr \int_0^\pi d\theta \, r \sin\theta (\cos 2 \theta+3) \nonumber \\
&=&\frac{4 \pi J  E_{\rm F}^2 q_{\parallel}^2}{3 \omega ^2}, \\
\label{eq:q_z}
\widetilde{\Pi}(q_z,\omega)
&\approx& \int_0^{E_{\rm F}} dr
\int_0^\pi d\theta
\int_0^{2\pi} d\phi
\mathcal{J}(r,\theta) \frac{2\Delta_{z}}{\omega^2}
\approx
\frac{4\pi q_z^2}{J\omega^2}\int_0^{E_{\rm F}} dr
\int_0^\pi d\theta \,  \left(r\sin\theta\right)^\frac{2-J}{J}r 
%\left(q_z \cos\theta+ \frac{\sin ^2\theta }{2 r} q_z^2\right) \nonumber \\
\left(\frac{\sin ^2\theta }{2 r}\right) \nonumber \\
&=&\frac{\pi^{\frac{3}{2}} \Gamma \left(1+\frac{1}{J}\right) E_{\rm F}^{\frac{2}{J}}}{\Gamma \left(\frac{3}{2}+\frac{1}{J}\right)}\frac{q_z^2}{\omega^2}.
\end{eqnarray}
\end{widetext}
Here we use the relation $\int_0^{\pi/2} d\theta \cos^m\theta \sin^n\theta=\frac{1}{2} B(\frac{m+1}{2},\frac{n+1}{2})$ where $B(m,n)=\frac{\Gamma(m)\Gamma(n)}{\Gamma(m+n)}$ is the beta function. Note that linear terms in $q_z$ vanish both in Eq.~(\ref{eq:q_parallel}) and Eq.~(\ref{eq:q_z}) due to the symmetry of the angular integrals. Combining Eq.~(\ref{eq:inter}) and the dielectric function in Eq.~(5) in the main text, we arrive at the analytic expressions for the leading-order long-wavelength plasma frequencies given in Eq.~(15) and Eq.~(16) in the main text.
%%%%%%%%%%%%%%%%%%%%%%%%%%%%%%%%%%%%%%%%%%%%%%%%%%

\end{document}